\documentclass[prc,twocolumn,showpacs,showkeys,nofootinbib,superscriptaddress]{revtex4}
\usepackage{graphicx}% Include figurefiles
\usepackage{dcolumn}
\usepackage{bm}

\newcommand{\svec}[1]{\bm{#1}}

\begin{document}

\topmargin -0.50in

\title{Anatomy of the $0\nu\beta\beta$ nuclear matrix elements}

\author{Fedor \v Simkovic}
%\email{fedor.simkovic@fmph.uniba.sk}
\altaffiliation{On  leave of absence from Department of Nuclear
Physics, Comenius University, Mlynsk\'a dolina F1, SK--842 15
Bratislava, Slovakia}
\affiliation{Institute f\"{u}r Theoretische Physik der Universit\"{a}t
T\"{u}bingen, D-72076 T\"{u}bingen, Germany}
\author{Amand Faessler}
%\email{amand.faessler@uni-tuebingen.de}
\affiliation{Institute f\"{u}r Theoretische Physik der Universit\"{a}t
T\"{u}bingen, D-72076 T\"{u}bingen, Germany}
\author{Vadim Rodin}
%\email{vadim.rodin@uni-tuebingen.de}
\affiliation{Institute f\"{u}r Theoretische Physik der Universit\"{a}t
T\"{u}bingen, D-72076 T\"{u}bingen, Germany}
\author{Petr Vogel}
%\email{pxv@caltech.edu}
\affiliation{Kellogg Radiation Laboratory and Physics Department, Caltech,
Pasadena, California, 91125, USA}
\author{Jonathan Engel}
%\email{engelj@physics.usc.edu}
\affiliation{Dept.\ of Physics and Astronomy, University of North Carolina,
Chapel Hill, NC 27599-3255, USA}

\begin{abstract}
We show that, within the Quasiparticle Random Phase Approximation (QRPA)
and the renormalized QRPA (RQRPA) based on the Bonn CD nucleon-nucleon
interaction, 
the competition between the pairing and the neutron-proton
particle-particle and particle-hole
interactions causes contributions to the
neutrinoless double-beta decay matrix element to nearly vanish
at internucleon distances of more than 2 or 3 fermis.  As a result,
the matrix element is more sensitive to short-range/high-momentum
physics than one naively expects. We analyze various ways of
treating that physics and quantify the uncertainty it produces in
the matrix elements, 
%which are obtained from a realistic effective nucleon-nucleon interaction, 
with three different treatments of short-range
correlations.
\end{abstract}



\date{\today}

\maketitle

\section{Introduction}

Neutrino oscillations are firmly established (see, e.g.
\cite{McKeown04,bil03,Langacker04,Strumia06,Fogli06}) and
demonstrate that neutrinos have masses many orders of magnitude
smaller than those of charged leptons. But since the masses are
nonzero, neutrinoless double-beta ($0\nu\beta\beta$) decay
experiments will likely tell us sooner or later whether neutrinos
are Majorana or Dirac particles \cite{FS98,V02,EV2002,EE04,AEE07}.
Moreover, the rate of the $0\nu\beta\beta$ decay, or limits on it,
can tell us about the absolute neutrino-mass scale and to some
extent about the neutrino mass hierarchy\footnote{Ref.\ \cite{matrix} 
discusses the goals and
future direction of the field.
Ref.\
\cite{nustudy} discusses issues particularly relevant for the
program of $0\nu\beta\beta$ decay search.}.  But to achieve these goals we need an accurate
evaluation of the nuclear matrix elements that govern the decay.

In this paper, which builds on previous publications \cite{Rod03a, Rod06}
(which we call I and II), we analyze some of the physics affecting the nuclear
matrix element $M^{0\nu}$ --- the competition between pairing and
neutron-proton particle-particle correlations, the nonintuitive dependence of
the decay amplitude on internucleon distance, and the treatment of short-range
correlations and other high-momentum phenomena --- that have not been
sufficiently discussed before. As in our earlier papers (and most attempts to
evaluate $M^{0\nu}$) we use the Quasiparticle Random Phase Approximation (QRPA)
and its generalization, the Renormalized QRPA (RQRPA), with an 
interaction obtained from the G matrix associated with the realistic Bonn CD
nucleon-nucleon interaction. That interaction, slightly renormalized, is used both
as the like particle pairing and as the neutron-proton force. 
Where appropriate, we compare the results to those
of the complementary Large-Scale Shell Model (LSSM).

The paper is organized as follows: In the next section, after
briefly summarizing the relevant formalism, we show that the final
value of $M^{0\nu}$ reflects two competing forces: the
like particle pairing
interaction that leads to the smearing of Fermi levels and the
residual neutron-proton interaction that, through ground state
correlations, admixes ``broken-pair" (higher-seniority) states. A
partial cancellation between these interactions increases the
sensitivity to their strengths. The same tendencies are present in
the LSSM, as a recent paper shows \cite{poves3}. 
(During the processing of this manuscript, a new paper on the
LSSM appeared \cite{SMII}, emphasizing the competition again.)
In section III we
discuss the dependence of $M^{0\nu}$  on the distance
between the two neutrons that are converted 
into two protons. 
We show that
the competition mentioned above implies that only internucleon
distances $r_{ij} \lesssim $ 2-3 fm contribute. That fact, not
recognized before, explains the sensitivity of the decay rate to 
higher order terms in nucleon currents,
nucleon form factors, and short-range nucleon-nucleon repulsion. We
show that the surprising dependence on  the internucleon distance
occurs not only in the QRPA but also in an
exactly solvable model \cite{eng04} that contains many ingredients
of real nuclear systems. Short-range correlations have recently
inspired a lively discussion \cite{Kort07a,Kort07b,Kort07c} and we
devote Section IV to various ways of treating them. In Section V we
present numerical results for nuclei of experimental interest, 
that include a comprehensive analysis,
within the QRPA method and its generalization, of the total uncertainty
of the $0\nu\beta\beta$ nuclear matrix elements, and compare with
results of the LSSM.  Section VI summarizes our findings.
Finally, in Appendix A we present formulae for two ways of evaluating 
the matrix elements, one via the evaluation of unsymmetrized two-body 
matrix elements (the procedure usually used) and another one
through the product of two one-body matrix elements. And in Appendix B
we show how to calculate shell-model particle-hole
decompositions so they can be compared with those calculated in the
QRPA.

\section{Formalism and multipole decompositions}

Throughout we assume that the $0\nu\beta\beta$ decay, if observed,
is caused by the exchange of the  Majorana neutrinos, the same
particles observed to oscillate. The half-life of the decay is then
\begin{equation}
\frac{1}{T_{1/2}} = G^{0\nu}(E_0,Z) |{M}^{0\nu}|^2
|\langle m_{\beta\beta} \rangle|^2~,
\end{equation}
where $G^{0\nu}(E_0,Z)$ is a precisely calculable phase-space factor
and ${M}^{0\nu}$ is the nuclear matrix element. The effective
Majorana neutrino mass $\langle m_{\beta\beta} \rangle$ is related
to the absolute mass scale and oscillation parameters through
\begin{equation}
\langle m_{\beta\beta} \rangle = \sum_i^N |U_{ei}|^2 e^{i\alpha_i} m_i ~,
~({\rm all~} m_i \ge 0)~,
\end{equation}
where $U_{ei}$ is the first row of the neutrino mixing matrix and
the and $\alpha_i$ are unknown Majorana phases. Any uncertainty in
${M}^{0\nu}$ makes the value of $\langle m_{\beta\beta} \rangle$
equally uncertain.

\begin{figure}[tb]
%   \epsfxsize=0.42\textwidth
 %   \epsffile{3rdpaper_fig1.eps}
%        \epsffile{3rdpaper_fig1.pdf}
\includegraphics[width=.42\textwidth]{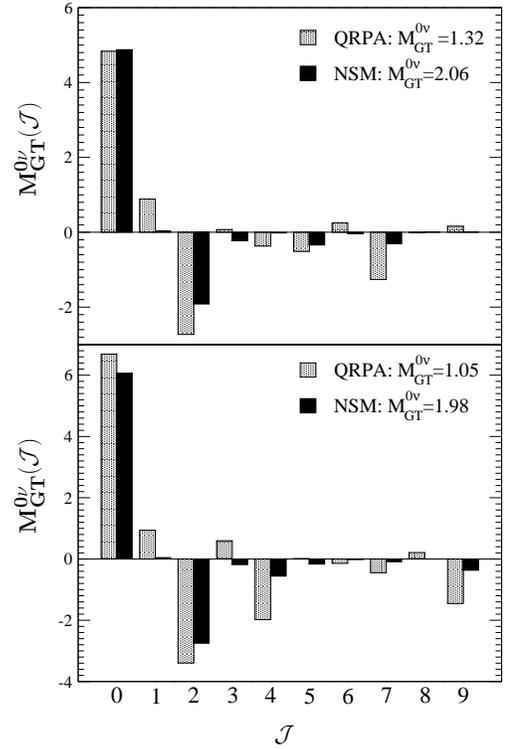}
%\vspace{0.3cm}
\caption{Contributions of different 
angular momenta
${\mathcal J}$ 
associated with the two decaying neutrons
to the Gamow-Teller part of $M^{0\nu}$ in $^{82}$Se (upper
panel) and $^{130}$Te (lower panel). The results of LSSM (dark
histogram) \cite{poves} and QRPA treatments (lighter histogram) are
compared. Both calculations use the same single-particle spaces:
($f_{5/2},p_{3/2},p_{1/2}, g_{9/2}$) for $^{82}$Se and ($g_{7/2},
d_{5/2}, d_{3/2}, s_{1/2}, h_{11/2}$) for $^{130}$Te. 
In the QRPA calculation the particle-particle interaction was
adjusted to reproduce the experimental $2\nu\beta\beta$-decay 
rate.} 
\label{fig:setesmJ}
\end{figure}

%Our phase space factors $G^{0\nu}(E_0,Z)$ are tabulated in Ref.\
%\cite{Sim99}. They agree quite closely with those given earlier in
%Ref. \cite{BV92}.  The $G^{0\nu}(E_0,Z)$ contain the inverse
%square of the nuclear radius $(R_{nucl})^{-2}$, compensated by the
%factor $R_{nucl}$ in ${M}^{0\nu}$. Different authors use different
%conventions for $R_{nucl}$ ($R_{nucl}$ = 1.2A$^{1/3}$ fm or
%1.1A$^{1/3}$ fm), a fact that is important to keep in mind when
%comparing the matrix elements without also looking at $G^{0\nu}(E_0,Z)$.

As stated above, we use the QRPA and RQRPA methods based
on the G matrix derived from the
realistic Bonn CD nucleon-nucleon force, i.e., the many body
hamiltonian is
\begin{equation}
H = \sum_{i=1}^A \frac{p_i^2}{2 m _p} + \frac{1}{2}   \sum_{i,j=1}^A V_{G-matrix}(i,j) ~.
\end{equation}
We describe in detail in Section V below the input used to solve the
corresponding well known equations of motion.

In the QRPA (and RQRPA) ${M}^{0\nu}$ is written as 
%sums 
a sum
over the virtual
intermediate states, labeled by their angular momentum and parity
$J^{\pi}$ and indices $k_i$ and $k_f$ (explanations of the notation
are in Appendix A, and II):
\begin{eqnarray}
\label{eq:long}
&&M_K  =  \sum_{J^{\pi},k_i,k_f,\mathcal{J}} \sum_{pnp'n'}
(-1)^{j_n + j_{p'} + J + {\mathcal J}} \times\qquad\qquad
\\
&&\sqrt{ 2 {\mathcal J} + 1}
\left\{
\begin{array}{c c c}
j_p & j_n & J  \\
 j_{n'} & j_{p'} & {\mathcal J}
\end{array}
\right\}  \times \qquad\qquad\qquad\qquad\qquad\quad
\nonumber \\
&&\langle p(1), p'(2); {\mathcal J} \parallel \bar{f}(r_{12})
%\tau_1^+ \tau_2^+
O_K \bar{f}(r_{12}) \parallel n(1), n'(2); {\mathcal J} \rangle \times
\qquad
\nonumber \\
&&\langle 0_f^+ ||
[ \widetilde{c_{p'}^+ \tilde{c}_{n'}}]_J || J^{\pi} k_f \rangle
\langle  J^{\pi} k_f |  J^{\pi} k_i \rangle
 \langle  J^{\pi} k_i|| [c_p^+ \tilde{c}_n]_J || 0_i^+ \rangle\, .
\nonumber 
\end{eqnarray}
The operators $O_K, K$ = Fermi (F), Gamow-Teller (GT), and Tensor
(T) contain neutrino potentials and spin and isospin operators, and
RPA energies $E^{k_i,k_f}_{J^\pi}$. The neutrino potentials, in
turn, are integrals over the exchanged momentum $q$,
\begin{eqnarray}
\label{eq:pot}
&&H_K (r_{12},E^k_{J^\pi}) =\qquad\qquad\qquad\qquad\qquad\\
&&\qquad\frac{2}{\pi g_A^2} {R} \int_0^{\infty}~ f_K(qr_{12})~
\frac{ h_K (q^2) q dq }
{q + E^k_{J^\pi} - (E_i + E_f)/2} \,.\nonumber
\end{eqnarray}
%\begin{equation}
%H_K (r_{12}) = \frac{2}{\pi g_A^2} {R} \int_0^{\infty}~ f_K(qr_{12})~
%\frac{ h_K (q^2) q dq }
%{q + E^m - (E_i + E_f)/2} ~,
%\label{eq:pot}
%\end{equation}
The functions $f_{F,GT}(qr_{12}) = j_0(qr_{12})$ and $f_{T}(qr_{12})
= j_2(qr_{12})$ are spherical Bessel functions (the sign of $j_2$ was given
incorrectly in Ref.\ \cite{Rod06}). The functions $h_K
(q^2)$ are defined in Appendix A and in II. The potentials depend
explicitly, though rather weakly, on the energies of the
virtual intermediate states, $E^k_{J^\pi}$. The function $\bar{f}(r_{12})$
in Eq. \ref{eq:long}
represents the effects of short range correlations.
These will be discussed in detail in Section IV.

%The reduced matrix elements of the one-body operators
%$c_p^+ \tilde{c}_n$ ($\tilde{c}_n$ denotes the time-reversed state)
%in the Eq. (\ref{eq:long})
%depend on the BCS coefficients $u_i,v_j$ and on the QRPA vectors
%$X,Y$ \cite{Sim99}. The difference between QRPA and RQRPA resides
%in the way these reduced matrix elements are calculated.

Two separate multipole decompositions are built into Eq.\
(\ref{eq:long}). One, already mentioned, is in terms the
$J^{\pi}$ of the virtual states in the intermediate nucleus, the
good quantum numbers of the QRPA and RQRPA. The other decomposition
is based on the angular momenta and parities ${\mathcal J}^{\pi}$ of the
pairs of neutrons that are transformed into protons with the same
${\mathcal J}^{\pi}$
% (We 
(we drop the superscript $\pi$ from now on for
convenience). This latter representation is particularly
revealing. In Fig. \ref{fig:setesmJ} we illustrate it both in the
LSSM and QRPA, with the same single-single particle spaces in each.
These two rather different approaches agree in a semiquantitative
way, but the LSSM entries for ${\mathcal J}
> 0$ are systematically smaller in absolute value.

Ref.\ \cite{poves3} makes the claim that QRPA results are too large because they
omit configurations with seniority greater than 4, which are especially
effective in canceling the pairing part of the matrix element.  This statement
is  not correct.  The QRPA does include configurations with higher
seniority (4,8,12, etc.) and, as the Fig.\ \ref{fig:setesmJ}  shows, the broken pair
contributions to the matrix elements are as large or larger than in the LSSM.
(Some of the difference might be due to differences in single-particle
energies and occupation numbers, which are not identical in the two
calculations even though the single-particle wave functions are.)
The reason that the QRPA results presented in Section V
are larger are a somewhat greater pairing
contribution and contributions from negative parity multipoles that reinforce
it.  Most of  the negative-parity contributions are absent from the shell model because
of restrictions on the model space.  These results suggest that the shell model
is as likely to be missing important physics as is the QRPA. We return to this
point in Section V.

\begin{figure}[tb]
%    \epsfxsize=0.45\textwidth
%    \epsffile{3rdpaper_fig2.eps}
%       \epsffile{3rdpaper_fig2.pdf}
\includegraphics[width=.35\textwidth,angle=-90]{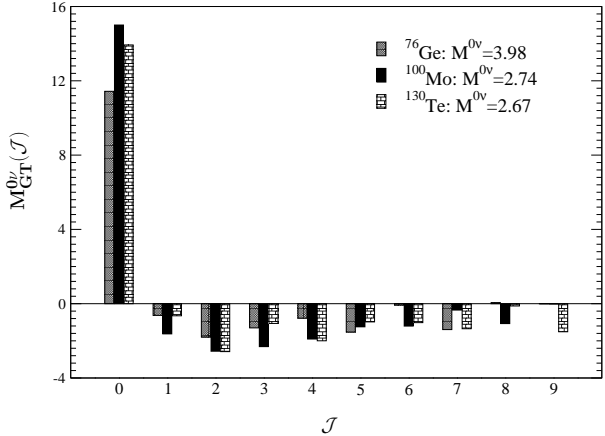}
%\includegraphics[width=.35\textwidth]{3rdpaper_fig2}
%\vspace{0.3cm}
\caption{Contributions of different 
angular momenta
${\mathcal J}$ 
associated with the two decaying neutrons
to $M^{0\nu}$ in
$^{76}$Ge, $^{100}$Mo and $^{130}$Te. We use the QRPA, with the
interaction strength $g_{pp}$ adjusted so that the $2\nu\beta\beta$
lifetime is correctly reproduced. Short-range correlations are
included the same way as in I and II.} \label{fig:gemote}
\end{figure}

In Fig.
\ref{fig:gemote} we show the ${\mathcal J}$ decomposition for three nuclei in
the QRPA, with single-particle spaces encompassing two major shells, a more
natural span for this method.  The cancellation between components with 
${\mathcal J} = 0$ 
pairs
and with ${\mathcal J} \ne 0$ 
pairs
is always pronounced. The net
${M}^{0\nu}$ is considerably smaller  than the pairing contribution and so
depends rather sensitively on the pairing physics that determines the
${\mathcal J} = 0$ part, as well as on the strength of the proton-neutron force
that determines the ${\mathcal J} \ne 0$ part.

From the structure of Eq.\ (\ref{eq:long}) and in particular from the
form of the reduced matrix elements $\langle 0_f^+ || [
\widetilde{c_{p'}^+ \tilde{c}_{n'}}]_J || J^{\pi} k_f \rangle$ and $
\langle  J^{\pi} k_i || [c_p^+ \tilde{c}_n]_J || 0_i^+ \rangle$ it
is obvious that only one of the two possible couplings between the
neutron and proton operators in the two-body matrix element $\langle
p(1), p'(2); {\mathcal J} \parallel \bar{f}(r_{12}) O_K \bar{f}(r_{12})
\parallel n(1), n'(2); {\mathcal J} \rangle$ is realized. This means
that this two-body matrix element should not be antisymmetrized. In the
LSSM one typically uses the closure approximation, which represents $M^{0\nu}$
as the ground-state-to-ground-state transition matrix element of a two-body operator. $M^{0\nu}$ can then be rewritten purely in terms of the antisymmetrized two-body matrix elements.
After antisymmetrization, however, it is not possible to recover the
decomposition into the multipoles $J^{\pi}$ of the virtual
intermediate states. In Appendix B we show how shell-model
practitioners, by retaining unsymmetrized matrix elements, can
decompose the matrix element into intermediate-state multipoles
$J^{\pi}$ for comparison with QRPA calculations.  We cannot,
however, make the comparison here without more shell-model data than
has been published.

\begin{figure}[tb]
%    \epsfxsize=0.4\textwidth
%    \epsffile{3rdpaper_fig3.eps}
%     \epsffile{3rdpaper_fig3.pdf}
\includegraphics[width=.4\textwidth]{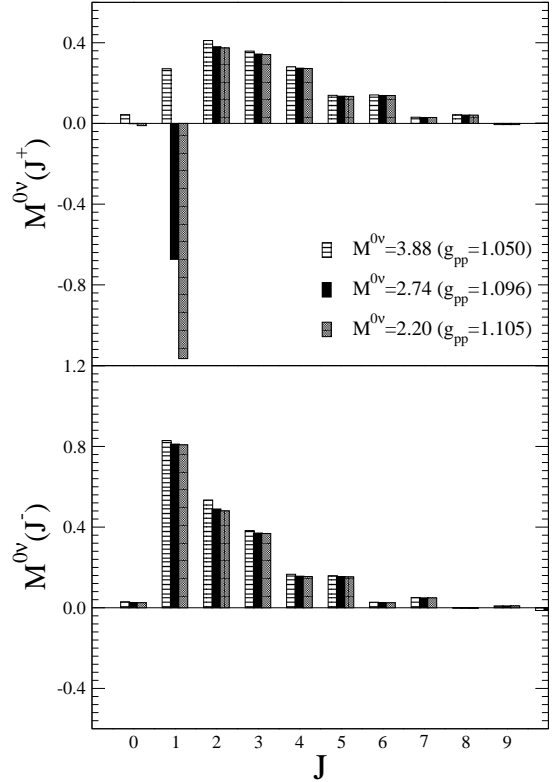}
%\vspace{0.3cm}
\caption{The contributions of different  
%multipoles $J$ %^{\pi}$
intermediate-state angular momenta $J$ 
to $M^{0\nu}$ in $^{100}$Mo (positive parities in the upper panel and negative
parities in the lower one). We show the results for several values of
$g_{pp}$. The contribution of the $1^+$ multipole changes
rapidly with $g_{pp}$, while those of the other
multipoles change slowly.}
\label{fig:mo100J}
\end{figure}

When using the QRPA or RQRPA to evaluate $M^{0\nu}$,
one must fix several important parameters, the effects of which  
were discussed in detail in I and II. The
strength $g_{pp}$ by which we renormalize the Bonn-CD G matrix in the
neutron-proton particle-particle channel
 is particularly important. We argued in I
and II that $g_{pp}$ should be chosen to reproduce the rate of
two-neutrino $\beta\beta$ decay.  This choice, among other things,
essentially removes the dependence of $M^{0\nu}$ on
the number of the single-particle states (or oscillator shells) in
the calculations. The $2\nu$ matrix element depends only on the $1^+$
multipole. In Fig. \ref{fig:mo100J} we show that it is essentially
this multipole that is responsible for the rapid variation of 
$M^{0\nu}$ with $g_{pp}$.  Fixing its contribution to a related
observable ($2\nu$ decay) involving the same initial and final
nuclear states appears to be an optimal procedure for determining
$g_{pp}$.

%In our numerical calculations we use the QRPA and RQRPA eigenvalues $E^m$ in the
%Eq. (\ref{eq:pot}) for each of the virtual intermediate states. However, as mentioned above
%the dependence of the final matrix element on the correct values of $E^m$ is relatively mild.
%Replacing  all of these  energies by a constant $0 < \langle E \rangle <$ MeV value
%results in the final matrix element to be within about 10\% of the value obtained with the correct $E^m$.
%Thus, we expect the error associated with the use of the closure approximation to
%be of similar magnitude.

\section{Dependence on the distance between the nucleons involved in the $0\nu\beta\beta$
transition.}

\begin{figure}[tb]
%    \epsfxsize=0.4\textwidth
%    \epsffile{3rdpaper_fig4.eps}
%    \epsffile{3rdpaper_fig4.pdf}
\includegraphics[width=.4\textwidth]{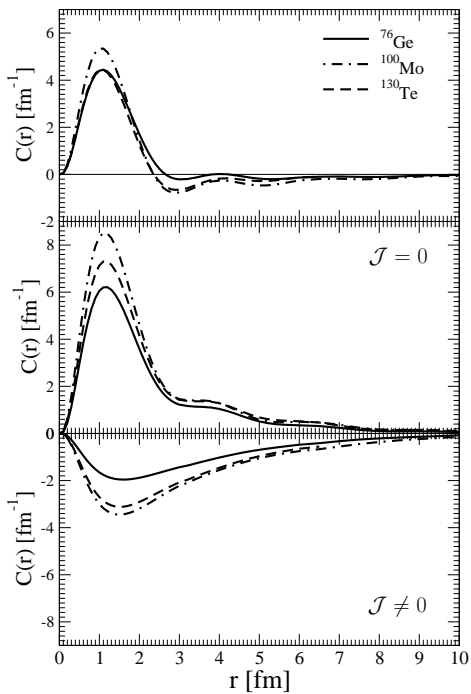}
%\vspace{0.3cm}
\caption{The dependence on $r_{12}$ of $M^{0\nu}$ for
$^{76}$Ge, $^{100}Mo$ and $^{130}$Te. The upper panel shows the full
matrix element, and the lower panel shows separately `pairing'
(${\mathcal J} = 0$ for the two decaying neutrons) 
and `broken pair' (${\mathcal J} \ne 0$)
contributions. The integrated matrix element is
$5.35$ for $^{76}$Ge, $4.46$ for $^{100}$Mo, and $4.09$ for
$^{130}$Te. The $g_{pp}$ values that reproduce the known
$T_{1/2}^{2\nu}$ are $1.030$, $1.096$ and $0.994$. The
single-particle space for $^{76}$Ge contains 9 levels (oscillator
shells $N=3,4$), and that for $^{100}$Mo and $^{130}$Te contains 13
levels (oscillator shells $N=3,4$ plus the $f$ and $h$ orbits from $N=5$).
Short-range correlation are not included, i.e. $\bar{f}(r_{12}) = 1$ in
Eq. (\ref{eq:long}).} \label{fig:r12_dep}
\end{figure}

The operators $O_K$ in Eq. (\ref{eq:long}) depend on the distance
$r_{12}$ between the two neutrons that are transformed into protons.
The corresponding neutrino potentials are the Fourier transforms
over the neutrino momentum $q$ as shown in Eq.(\ref{eq:pot}).
Obviously, the range of $r_{12}$ is restricted from above by $r_{12}
\le 2R_{nucl}$. We show here, however, that in reality only much
smaller values, $r_{12} \lesssim $ 2-3 fm, or equivalently larger
values of $q$, are relevant. Thus a good
description of the physics involving distances $r_{12} \sim$ 1 fm,
or $q \sim$ 200 MeV is important. That finding has not
been recognized before, but perhaps it should be not so surprising
that $q \sim p_{Fermi}$ is the most relevant momentum transfer.

\begin{figure}[tb]
%    \epsfxsize=0.4\textwidth
%    \epsffile{3rdpaper_fig5.eps}
%        \epsffile{3rdpaper_fig5.pdf}
\includegraphics[width=.4\textwidth]{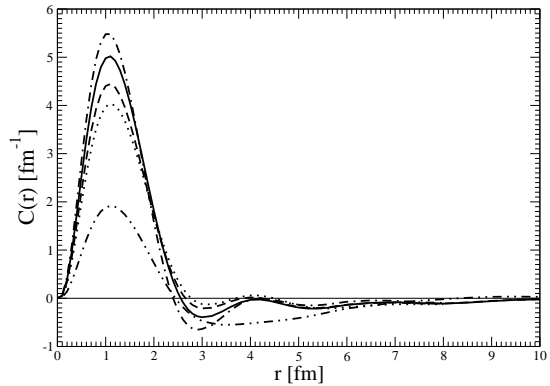}
%\vspace{0.3cm}
\caption{The $r_{12}$ dependence of $M^{0\nu}$ for
$^{76}$Ge, from calculations with different number of
single-particle orbits. The dot-dashed curve was obtained with 21
s.p. subshells, the full curve with 12 subshells, the dashed curve 
with 9 subshells, the dotted curve with 6 subshells and the double
dot-dashed curve with only 4 subshells.  } \label{fig:msd}
\end{figure}

An example of the $r_{12}$ dependence of $M^{0\nu}$ is shown in Fig.
\ref{fig:r12_dep} for three nuclei.  The quantity $C(r)$ is defined
by evaluating $M^{0\nu}$ after multiplying $H^K(r',E^k_{J^\pi})$ by
$r^2 \delta(r-r')$, so that $C(r)$ is the contribution at $r$ to
$M^{0\nu}$, with $\int_0^\infty C(r) dr = M^{0\nu}$. As the lower
panel of the figure demonstrates, the cancellation between the
${\mathcal J} = 0$ and ${\mathcal J} \ne 0$ components is
essentially complete for $r_{12} \gtrsim$ 2-3 fm.  Since the typical
distance from a particular nucleon to its nearest neighbor is $\sim$
1.7 fm (because $R_{nucl}$ = 1.2A$^{1/3}$) the nucleons
participating in the $0\nu\beta\beta$ decay are mostly nearest
neighbors. Short-range nucleon-nucleon repulsion, the finite nucleon
size, represented by nucleon form factors, and components of the
weak currents that are typically suppressed by $q/M_{nucleon}$ are
therefore more important than one would naively expect.

\begin{figure}[tb]
%    \epsfxsize=0.4\textwidth
%    \epsffile{3rdpaper_fig6.eps}
%        \epsffile{3rdpaper_fig6.pdf}
\includegraphics[width=.4\textwidth]{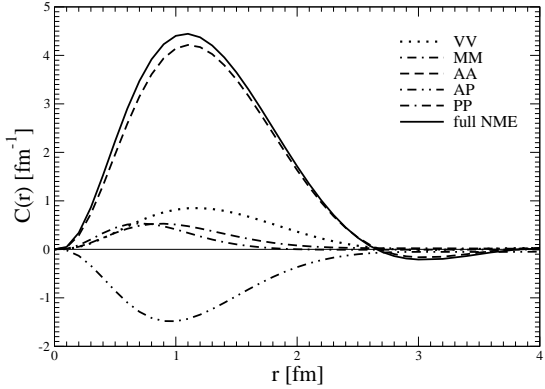}
%\vspace{0.3cm}
\caption{The $r_{12}$ dependence of the contributions of various pieces of $M^{0\nu}$ for $^{76}$Ge are shown. Here AA stands for axial, VV for vector, AP  for axial-pseudoscalar interference, PP for
pseudoscalar, and MM for weak-magnetism. For definitions see Appendix A, or
Ref.\ \cite{Sim99}. The model space contains 9 subshells.}
\label{fig:hot}
\end{figure}

Perhaps the most interesting thing about the figure is that the
pairing and non-pairing parts of $C(r)$ taken individually (as in
the two panels of the figure) extend to significantly larger $r$.
The cancellation between them, that we discussed earlier, is particularly
effective beyond 2 or 3 fm, leaving essentially nothing there.
Figure \ref{fig:msd} shows that the shape of $C(r)$, like the
integrated matrix element, is essentially independent of the number of 
single-particle orbits included, as long as the truncation is not too
severe (as it is with the dash-double-dot curve, for which important
spin-orbit partners were omitted -- only the 4 single particle
states $p_{3/2}, p_{1/2}, f_{5/2}, g_{9/2}$ were included) and the
coupling constant $g_{pp}$ is chosen to reproduce the measured
$2\nu\beta\beta$ lifetime.   For other values of $g_{pp}$ the
cancellation between the ${\mathcal J} = 0$ and ${\mathcal J} \ne 0$
contributions at $r$ larger than 2 or 3 fm is not as complete as in Fig.
\ref{fig:r12_dep}.  We return to this point shortly

%We demonstrate that the behavior of the $r_{12}$ dependence, namely
%the cancellation of the contribution from $r_{12} \gtrsim$ 2-3 fm is
%independent on   the number of the single-particle orbits in
%Fig.\ref{fig:msd}. All curves in that figure were obtained with the
%parameter  $g_{pp}$ chosen in each case in such a way that the
%$2\nu\beta\beta$-decay rate is correctly reproduced. The resulting
%$M^{0\nu}$ matrix elements are almost constant, varying between 5.0 - 6.0 . The
%dash-double dot curve, obtained with only 4 single particle states
%($p_{3/2}, p_{1/2}, f_{5/2}, g_{9/2}$), i.e. without the spin-orbit
%partners of $f_{5/2}$ and $g_{9/2}$ orbits show a different
%behavior; the larger $r_{12}$ contribute noticeably, and with the
%opposite sign from the contribution of smaller $r_{12}$ values. The
%corresponding  $M^{0\nu}$ are also much smaller  in that case,
%despite the adjustment of $g_{pp}$ that acquires unrealistically
%large value of ???.

We show the $r_{12}$ dependence of the different parts of the
$M^{0\nu}$ in Fig.\ \ref{fig:hot}. All individual
contributions die out at $r$ larger than 2 or 3 fm.  The
pseudoscalar-axial vector interference part has opposite sign from
the other contributions, and essentially (and accidentally) cancels
the contributions of the vector, weak magnetism and pure
pseudoscalar pieces. 
The higher-order terms reduce the matrix element noticeably, and have to
be included.

\begin{figure}[tb]
 %       \epsfxsize=0.42\textwidth
%    \epsffile{3rdpaper_fig7.eps}
%        \epsffile{3rdpaper_fig7.pdf}
\includegraphics[width=.42\textwidth]{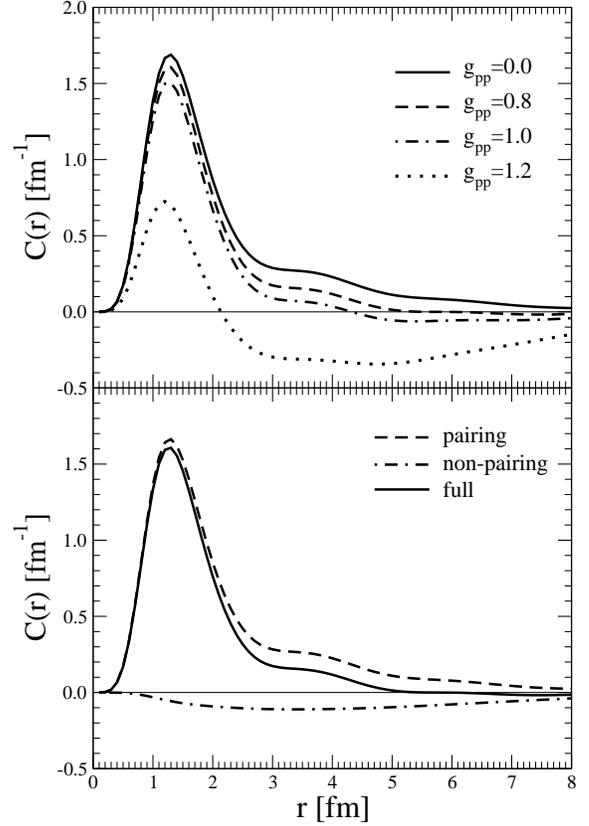}
%\vspace{0.3cm}
\caption{\label{fig:smodel} The $r_{12}$ dependence of $M^{0\nu}$ in
the exactly solvable model for 4 values of $g_{pp}$ (upper panel).
The integrated matrix elements are 2.93 for $g_{pp} = 0$ and 1.69 for $g_{pp}
= 1$.  The lower panel shows separately the contributions of ${\mathcal J} =0$ 
pairs
and ${\mathcal J} \ne 0$ 
pairs
for  $g_{pp} = 0.8$. }
\label{fig:model1}
\end{figure}

To gain some insight into the renormalization of the double-beta
decay operator in the shell model, Ref.\ \cite{eng04}  employs a
solvable model based on the algebra $SO(5)\times SO(5)$. The valence
space contains two major shells ($fpg_{9/2}$ and $sdg_{7/2}$), split
by an energy $\epsilon$, with degenerate levels within each shell.
The single-particle wave functions are taken from a harmonic
oscillator with $\hbar\omega=$9.2 MeV.  The Hamiltonian (for this
schematic model only, not in the rest of the paper) is

%\vspace{.2in}
\begin{eqnarray}
\label{eq:H} H=\epsilon
\hat{N}_2\hspace{7cm}&&  \\
-G   \sum_{a,b=1}^2 \left( S^{\dag a}_{pp} S^b_{pp} + S^{\dag
a}_{nn} S^b_{nn} + g_{pp} S^{\dag a}_{pn} S^b_{pn}
  - g_{ph}
\svec{T}_a \cdot \svec{T}_b \right) ~,&& \nonumber
\end{eqnarray}
where $a,b=1,2$ label the shells (lower and upper), $\epsilon$ is
the energy difference between the shells, $\hat{N}_2$ is the number
operator for the upper shell, $\svec{T}_a$ is total isospin operator
for shell $a$, and
\begin{eqnarray}
\label{eq:S} S^{\dag a}_{pp} &= &\frac{1}{2}\sum_{\alpha \in a}
\hat{j}_{\alpha}
[\pi^{\dag}_{\alpha} \pi^{\dag}_{\alpha}]^0_0 \nonumber \\
S^{\dag a}_{nn} &= &\frac{1}{2}\sum_{\alpha \in a} \hat{j}_{\alpha}
[\nu^{\dag}_{\alpha} \nu^{\dag}_{\alpha}]^0_0 \nonumber \\
S^{\dag a}_{pn} &= &\frac{1}{\sqrt{2}}\sum_{\alpha \in a}
\hat{j}_{\alpha}[\pi^{\dag}_{\alpha} \nu^{\dag}_{\alpha}]^0_0  ~.
\end{eqnarray}
Here $\pi^{\dag}_{\alpha}$ ($\nu^{\dag}_{\alpha}$) creates a proton
(neutron) in level $\alpha$ with angular momentum $j_{\alpha}$,
$\hat{j} \equiv \sqrt{2j+1}$, and the square brackets indicate
angular-momentum coupling.  $H$ contains only generators of $SO(5)
\times SO(5)$, so its lowest lying eigenstates consist of
configurations in which the nucleons are entirely bound in isovector
$S$ pairs of the type in Eq.\ (\ref{eq:S}).

This model has no active spin, so it is only suitable for
calculating Fermi (neutrinoless) double-beta decay.  To simulate the
effect of $g_{pp}$ on Gamow-Teller decay we change the Fermi matrix
element by varying the strength of isovector neutron-proton
pairing, in the same way that we change the Gamow-Teller matrix
element in the realistic QRPA by varying the strength of isoscalar
pairing. The advantage of this model is that we can solve it exactly
rather than in the QRPA.

\begin{figure}[tb]
%    \epsfxsize=0.45\textwidth
%    \epsffile{3rdpaper_fig9.eps}
 %       \epsffile{3rdpaper_fig9.pdf}
 \includegraphics[width=.45\textwidth]{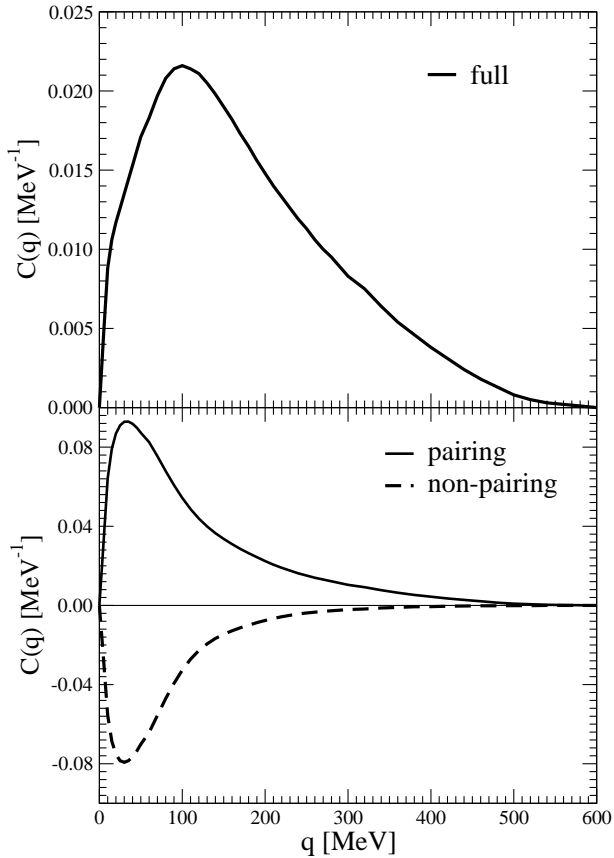}
%\vspace{0.3cm}
\caption{The momentum-transfer dependence of $M^{0\nu}$ in $^{76}$Ge. 
The
upper panel is for the full matrix element; in the lower panel we separate the
${\mathcal J} = 0$ and ${\mathcal J} \ne 0$ parts. The scale is
different in the two panels.  
The model space contains 9 subshells.} 
\label{fig:qdep}
\end{figure}

We can use the model to test the $\svec{r}_{12}$-dependence of the
double-beta decay matrix element in an exact solution. (Analytic
expressions for the necessary matrix elements are in Ref.\
\cite{eng04}.) The upper panel of Fig.\ \ref{fig:smodel} shows the
dependence for several values of $g_{pp}$, with $g_{ph}=0$ and
$\epsilon=10 G$.  Just as in the realistic QRPA calculations, the
contribution beyond $r=3$ fm is very small for $g_{pp}$ around 1; it
is too small to distinguish from zero in the figure beyond 5 fm for
$g_{pp}=0.8$.  However, for other values, as noted above, the
large-$r$ contributions can be substantial.  The bottom panel
divides the function into like-particle pairing and non-pairing parts for
$g_{pp}=0.8$. The two cancel to high precision at large $r$. The
suppression at long ranges we observe in the QRPA, then, appears to
be fairly general. It happens even in a very simple model, solved
exactly.(As noted above, a new preprint \cite{SMIII} appeared during
the processing of this manuscript.
In it the $r_{12}$ dependence
of the $M^{0\nu}$, as well as the dependence of the separated pairing and
broken-pair contributions,
was evaluated in the LSSM. That analysis, inspired by our work, yielded
curves that are
strikingly similar to those in Fig.\ \ref{fig:r12_dep}.)

\begin{figure}[tb]
%   \epsfxsize=0.45\textwidth
%    \epsffile{3rdpaper_fig8.eps}
 %       \epsffile{3rdpaper_fig8.pdf}
 \includegraphics[width=.45\textwidth]{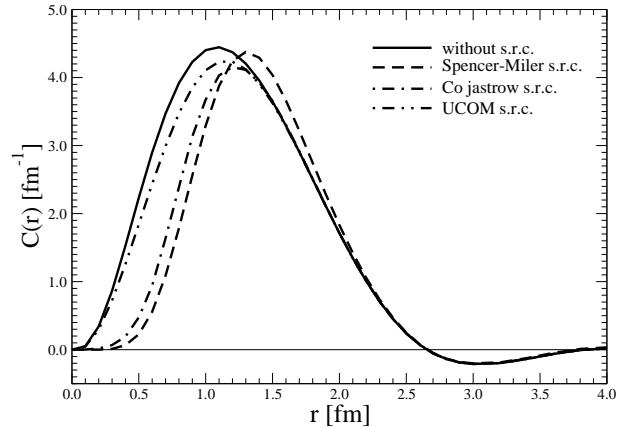}
%\vspace{0.3cm}
\caption{The $r_{12}$ dependence of $M^{0\nu}$ in
$^{76}$Ge evaluated in the model space that contains 9 subshells. 
The four curves show the effects of different treatments
of short-range correlations. The resulting $M^{0\nu}$ values are
5.32 when the effect of short range correlations
is ignored, 5.01 when the UCOM transformation
\cite{UCOM} is applied, 4.14 when the $\bar{f}(r_{12})$ from
Fermi hypernetted-chain calculations \cite{Co} is used in Eq.(\ref{eq:long}), and 3.98 when
the phenomenological Jastrow $\bar{f}(r_{12})$ is used \cite{Spencer}.  
} \label{fig:src}
\end{figure}

\section{Short-range correlations and other high-momentum phenomena}

Since only $r_{12} \lesssim  2-3$ fm, (equivalently $q > \hbar c/(2
- 3{~\rm fm})$), contributes to $M^{0\nu}$, some
otherwise negligible effects become important. These effects are not
commonly included, or included only in rough approximation, in
nuclear-structure calculations. For example, the dipole
approximation for nucleon form factors and the corresponding
parameters $M_V$ and $M_A$ come from electron-
and neutrino charged-current-scattering from on-shell nucleons.
Nuclear structure deals with bound nucleons and virtual neutrinos
that are far off-shell. Similarly, the induced pseudoscalar current,
with its strength obtained from the Goldberger-Treiman relation, has
been tested in muon capture on simple systems. Here we are using
this current for off-shell
virtual neutrinos. Short range nucleon-nucleon repulsion has been
considered carefully when calculating nuclear binding energies, but
here we need its effect on a transition operator connecting two
different nuclear ground states. All these effects will introduce
some uncertainty because their treatment is not well tested.
Nevertheless, it is important to understand their size at least
roughly.

\begin{figure}[tb]
%    \epsfxsize=0.4\textwidth
%    \epsffile{3rdpaper_fig10.eps}
%        \epsffile{3rdpaper_fig10.pdf}
\includegraphics[width=.4\textwidth]{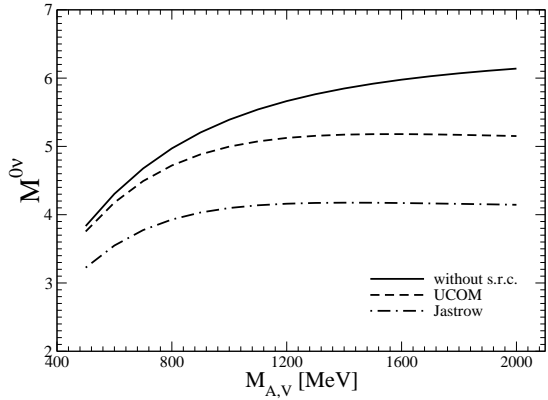}
%\vspace{0.3cm}
\caption{The dependence of $M^{0\nu}$ in $^{76}$Ge on the value of the
dipole-form-factor cut-off parameters $M_{A,V}$. The upper curve was
calculated without short range correlations, while the two lower
curves were obtained with the UCOM and Jastrow methods.  The two
lower curves, unlike the upper one, are essentially flat for
$M_{A,V}$ larger than the standard value of $\approx 1$ GeV.
} \label{fig:formf}
\end{figure}

To show the importance of high momenta explicitly, we display in
Fig.\ \ref{fig:qdep} the $q$ dependence $C(q)$ --- defined in complete analogy
to $C(r)$ --- of $M^{0\nu}$
in $^{76}$Ge, in a similar manner to which we exhibited the $r_{12}$
dependence earlier. The cancellation between
the ${\mathcal J} = 0$ and ${\mathcal J} \ne 0$ parts is
particularly complete at lower values of $q$ so that the resulting curve 
in the upper panel,
although reduced in magnitude, is clearly shifted towards higher
$q$.
%The figure represents the
%complementary picture to the $r_{12}$ dependence shown in Fig.
%\ref{fig:r12_dep}.

The first high-momentum effect we examine is short-range
correlations. Fig.\ \ref{fig:src} displays the
$r_{12}$ dependence of $M^{0\nu}$ for several methods of handling
short-range physics. For obvious reasons all methods reduce the
magnitude of $M^{0\nu}$. The Unitary Correlation Operator Method
(UCOM) \cite{UCOM} leads to the smallest reduction, less
than 5\%. The phenomenological Jastrow-like function $\bar{f}(r_{12})$ in
Eq.(\ref{eq:long}) (from Ref.\ \cite{Spencer}) reduces $M^{0\nu}$
by about 20\%. We also display the results of using a
microscopically-derived Jastrow function \cite{Co}; its effect is similar to that of the phenomenological function.
Since it is not clear which approach is best, we believe it prudent
to treat the differences as a relatively modest
uncertainty.

Nucleon form factors pose fewer problems because it turns out that
once the short-range correlations effects are included, no matter
how, the form factors are almost irrelevant as long as the cut-off masses
$M_{A,V}$ are at least as large as the standard values 
($M_{A} = 1.09$ GeV and $M_V=0.85$ GeV). In Fig. \ref{fig:formf} we
show the dependence of $M^{0\nu}$ on the values of
$M_{A,V}$ which for this purpose are set equal to each other, with
three alternatives for treating short-range correlations. By 2 GeV
the curves have essentially reached the infinite-mass limit. Since
they are essentially flat past 1 GeV for both the UCOM and
Jastrow-like prescriptions, including the form factors causes only
minor changes in $M^{0\nu}$.  Only if the correlations are ignored
altogether do the form factors make a significant difference.

Finally, there is little doubt that the higher order weak currents,
induced pseudoscalar and weak magnetism, should be included in the
calculation. Even though the Goldberger-Treiman relation has not
been tested in two-body operators, the relation is sufficiently well
established that we do not associate a sizable uncertainty with its
use.

\section{Numerical results}

\begin{figure}[tb]
%    \epsfxsize=0.48\textwidth
%    \epsffile{3rdpaper_fig11.eps}
%        \epsffile{3rdpaper_fig11-1.pdf}
\includegraphics[width=.48\textwidth]{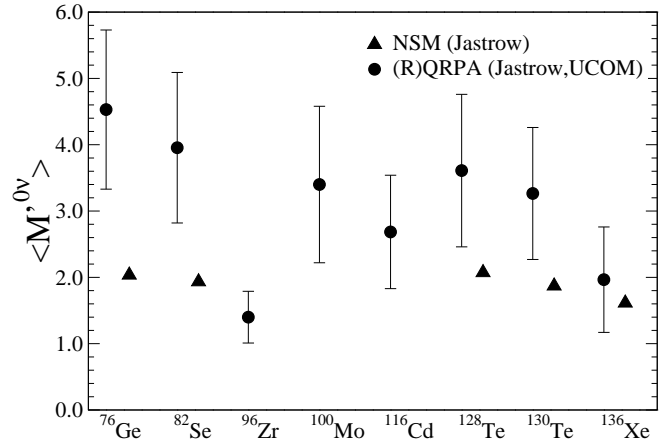}
%\vspace{0.3cm}
\caption{
Circles represent the mean value of the upper and lower 
limits of our calculated values of $M^{'0\nu}$; 
see text for description of error bars.  For comparison
the results of a recent Large Scale Shell Model evaluation of
$M^{'0\nu}$ that used the Jastrow-type treatment of short range
correlations are also shown as triangles}.  
\label{fig:total}
\end{figure}

Fig.\ \ref{fig:total} shows our calculated ranges for $M'^{0\nu}$,
defined as $M'^{0\nu}=(g_A/1.25)^2 M^{0\nu}$ to allow us to display the effects
of uncertainties in $g_A$. 
Such a definition allows us to use the same phase space factor $G^{0\nu}$
with $g_A = 1.25$ when calculating the $0\nu\beta\beta$-decay rate. 

We consider the effects of
short-range correlations an uncertainty, since we don't know the best way
of treating them.  The error bars in Fig. \ref{fig:total}
represent the difference between the highest
and lowest of 24 calculations --- in either the QRPA or RQRPA, each with  3 different sets
of single-particle states
(usually 2, 3 and 4 oscillator shells), 
2 values for the in-medium $g_A$ (1.0 and
1.25), and 2 treatments of short-range correlations (phenomeological
Jastrow functions and the UCOM method) --- and include the experimental
uncertainty in the values of the $2\nu$ lifetimes used to determine $g_{pp}$.
Thus the error bars displayed in Fig. \ref{fig:total} represent our estimate
of the full uncertainty in the $0\nu\beta\beta$ matrix elements within the
QRPA and RQRPA methods.
Though the results are in reasonable
agreement with those presented in Refs.\
\cite{Kort07b,Kort07c}, the uncertainties are different.
%For comparison
%the $M^{0\nu}$ values from the most recent QRPA calculation in Ref. \cite{Kort07c}
%are also shown. The agreement between the two QRPA based approaches is
%quite promising.

Reference \cite{poves2} presents new LSSM results for $M'^{0\nu}$,
the values of which are shown also in Fig. \ref{fig:total}.  They are
somewhat smaller than the QRPA values for $^{76}$Ge and $^{82}$Se
and in a fair agreement for $^{116}$Cd, $^{128}$Te, $^{130}$Te and
$^{136}$Xe. 
%The LSSM calculation does not include nucleon currents
%beyond the allowed approximation, but other
Various nuclear-structure
effects may be responsible for the discrepancies. More complicated
configurations in LSSM that are absent in QRPA typically reduce 
$M^{0\nu}$ and could make most of the difference. On
the other hand, in the QRPA includes more single particle states
than in LSSM. That has the tendency to increase $M^{0\nu}$ 
and could also be responsible for the discrepancy.  As we said earlier, the
notion that the QRPA omits high-seniority states is not correct and
shouldn't be used to argue that the LSSM results are more accurate.

Given the interest in the subject, we show the range of predicted half-lives
corresponding to our full range of $M'^{0\nu}$ in Table \ref{tab:t12}
(for $\langle m_{\beta\beta} \rangle$ = 50 meV).
As we argued above, this is a rather conservative range within
the QRPA and its related frameworks. One should keep in mind, however,
the discrepancy between the QRPA and LSSM results as well as systematic
effects that might elude either or both calculations.

\begin{table*}[htb]
  \begin{center}
\caption{\label{tab:t12}The calculated ranges of the nuclear matrix element
$M^{'0\nu}$ evaluated within both the QRPA and RQRPA and with both standard
($g_A = 1.254$) and quenched ($g_A = 1.0$) axial-vector couplings.  In each
case we adjusted  $g_{pp}$ so that the rate of the $2\nu\beta\beta$=decay is
reproduced.  Column 2 contains the ranges of $M^{'0\nu}$ with the
phenomenological Jastrow-type treatment of short range correlations (see I and
II), while column 3 shows the UCOM-based results (see Ref. \cite{UCOM}).
Columns 3 and 5 give the  $0\nu\beta\beta$-decay half-life ranges corresponding
to the matrix-element ranges in columns 2 and 4, for
$<m_{\beta\beta}>=50$~meV.  }

\vspace{.2cm}

\begin{tabular}{lccccc}
\hline\hline
 Nuclear & \multicolumn{2}{c}{(R)QRPA (Jastrow s.r.c.)} & &
           \multicolumn{2}{c}{(R)QRPA (UCOM s.r.c.)}\\ \cline{2-3} \cline{5-6}
transition & $M^{'0\nu}$ & $T^{0\nu}_{1/2}$ ($\langle m_{\beta\beta} \rangle$ = 50 meV) &
           & $M^{'0\nu}$ & $T^{0\nu}_{1/2}$ ($\langle m_{\beta\beta} \rangle$ = 50 meV) \\\hline
$^{76}Ge\rightarrow {^{76}Se}$
  &  $(3.33,4.68)$ & $(6.01,11.9)\times 10^{26}$ &
  &  $(3.92,5.73)$ & $(4.01,8.57)\times 10^{26}$ \\
$^{82}Se\rightarrow {^{82}Kr}$
  &  $(2.82,4.17)$ & $(1.71,3.73)\times 10^{26}$ &
  &  $(3.35,5.09)$ & $(1.14,2.64)\times 10^{26}$ \\
$^{96}Zr\rightarrow {^{96}Mo}$
  &  $(1.01,1.34)$ & $(7.90,13.9)\times 10^{26}$ &
  &  $(1.31,1.79)$ & $(4.43,8.27)\times 10^{26}$ \\
$^{100}Mo\rightarrow {^{100}Ru}$
  &  $(2.22,3.53)$ & $(1.46,3.70)\times 10^{26}$ &
  &  $(2.77,4.58)$ & $(8.69,23.8)\times 10^{25}$ \\
$^{116}Cd\rightarrow {^{116}Sn}$
  &  $(1.83,2.93)$ & $(1.95,5.01)\times 10^{26}$ &
  &  $(2.18,3.54)$ & $(1.34,3.53)\times 10^{26}$ \\
$^{128}Te\rightarrow {^{128}Xe}$
  &  $(2.46,3.77)$ & $(3.33,7.81)\times 10^{27}$ &
  &  $(3.06,4.76)$ & $(2.09,5.05)\times 10^{27}$ \\
$^{130}Te\rightarrow {^{130}Xe}$
  &  $(2.27,3.38)$ & $(1.65,3.66)\times 10^{26}$ &
  &  $(2.84,4.26)$ & $(1.04,2.34)\times 10^{26}$ \\
$^{136}Xe\rightarrow {^{136}Ba}$
  &  $(1.17,2.22)$ & $(3.59,12.9)\times 10^{26}$ &
  &  $(1.49,2.76)$ & $(2.32,7.96)\times 10^{26}$ \\
\hline\hline
\end{tabular}
  \end{center}
\end{table*}

\section{Conclusions} 

The most important and novel result here is that 
the generic competition between 
${\mathcal J = 0}$ (pairing) and ${\mathcal J \ne 0} $ (broken
pair) multipoles leads to almost complete cancellation of the
contribution to the matrix element from internucleon distances $r \gtrsim$ 2-3 fm. That
explains why the effects that depend on smaller values of $r$, or
equivalently larger momentum transfers $q$, become important.
This competition also means that the final matrix elements have
enhanced sensitivity to the strengths of these interactions.
Despite the uncertainties associated with the short range effects,
we conclude that a proper fitting of the QRPA and/or RQRPA parameters
leads to a relatively narrow range for
$M'^{0\nu}$, with a smooth  dependence ($^{96}$Zr being an exception)
on atomic charge and mass.

We evaluate the values of the matrix elements
for nuclei of experimental interest and display our best estimate of the
corresponding spread. Part of that spread is associated with the difference
in the size of the single particle space and whether QRPA or RQRPA
is used, as discussed earlier in I and II.
An interesting new conclusion is that short-range correlations, no matter how
they are treated, essentially eliminate the effect of finite nucleon size on
the matrix elements.  But we still do not know the best way to treat the
correlations, a fact that contributes about 20\% to uncertainties presented
above.  The uncertainty in the effective value of $g_A$ contributes about
30\%, with the rest due to choice of method and model space, and the
experimental uncertainty in $2\nu$ lifetimes.

\acknowledgments

F.\v{S}, V.R. and A.F. acknowledge the support of  the EU ILIAS project under the contract
RII3-CT-2004-506222, the Transregio Project TR27 "Neutrinos and Beyond",
the Deutsche Forschungsgemeinschaft (436 SLK 17/298) and, in addition, 
F.\v{S} was supported by  the VEGA Grant agency of
the Slovak Republic under the contract No.~1/0249/03. The work of J.E. was
partially supported by the U.S.\ Department of Energy under Contract DE-FG02-97ER41019
Two of us (P.V. and J.E.) thank the INT at the University of Washington
for the hospitality during the initial stage of this work and the US Department
of Energy for partial support.

\appendix
\section{}
Here we outline the derivation of $M^{0\nu}$,
emphasizing
induced (higher-order) currents.
We assume light-neutrino exchange throughout, and the standard lepton$\times$hadron weak
charged-current Hamiltonian.
The hadronic current, expressed in terms of nucleon fields $\Psi$, is 
\begin{eqnarray}
j^{\rho \dagger}
&=&  \overline{\Psi} \tau^+ \left[ g_V(q^2) \gamma^\rho
+ i g_M (q^2) \frac{\sigma^{\rho \nu}}{2 m_p} q_\nu \right.
\nonumber\\
&&~~~~~~~~~\left.
 - g_A(q^2) \gamma^\rho\gamma_5 - g_P(q^2) q^\rho \gamma_5 \right] \Psi,
\label{a1}
\end{eqnarray}
where $m_p$ is the nucleon mass and $q^\mu$ is the momentum transfer,
i.e.\ the momentum of the virtual neutrino.
Since in the $0\nu\beta\beta$ decay $\vec{q}^2 \gg q_0^2$ we take
$q^2 \simeq - \vec{q}^2$.
For the vector and axial vector form factors we adopt the usual dipole approximation
$g_V({\vec q}^{~2}) = {g_V}/{(1+{\vec q}^{~2}/{M_V^2})^2}$,
$g_A({\vec q}^{~2}) = {g_A}/{(1+{\vec q}^{~2}/{M_A^2})^2}$,
with $g_V$ = 1, $g_A$ = 1.254, $M_V$ = 850 MeV, and $M_A$ = 1086 MeV.
We use the usual form for weak magnetism,
$g_M({\vec q}^{~2})= (\mu_p-\mu_n) g_V({\vec q}^{~2})$,
and the Goldberger-Treiman relation, $g_P({\vec q}^{~2}) = {2 m_p g_A({\vec
q}^{~2})}/({{\vec q}^{~2} + m^2_\pi})$, for the induced pseudoscalar term.

To derive the expression for the matrix element we follow the procedure outlined
in Ref. \cite{bilpet}, arriving after a few steps at an expression for the $0_i^+ \rightarrow 0_f^+$
ground state to ground state transition:
\begin{eqnarray}
M^{0\nu}= \frac{4\pi R}{g_A^2}\int
\left(\frac{1}{(2\pi)^3}\int \frac{e^{-i \vec{q}.(\vec{x}_1-\vec{x}_2)}}{|q|}\right)
\times \qquad\qquad \\
\sum_m \frac{<0^+_f|J_{\alpha}^{\dagger}(\vec{x}_1)|m><m| J^{\alpha\dagger}(\vec{x}_2)|0^+_i>}
{E_m - (E_i + E_f)/2 + |q|}
d\vec{q} d\vec{x}_1 d\vec{x}_2\,.\nonumber
\label{a2}
\end{eqnarray}
We have made the (accurate) approximation that all electrons are emitted in the $s_{1/2}$ state,
with energies equal to
half the available energy $(E_i-E_f)/2$.
The normalization factor ${4\pi~R}/{g_A^2}$, introduced for convenience, is
compensated for by corresponding
factors in the phase space integral.

Reducing the nucleon
current to the non-relativistic form yields (see Ref.\cite{erwe})
in Eq.(\ref{a2}):
\begin{equation}
J^{\rho\dagger}(\vec{x})=\sum_{n=1}^A \tau^+_n [g^{\rho 0} J^0({\vec q}^{~2}) +
\sum_k g^{\rho k}  J^k_n({\vec q}^{~2})] \delta(\vec{x}-{\vec{r}}_n),
\label{a3}
\end{equation}
where $J^0({\vec q}^{~2}) = g_V(q^2)$ and
\begin{equation}
{\vec J}_n({\vec q}^{~2}) =  g_M({\vec q}^{~2})
i \frac{{\vec{\sigma}}_n \times \vec{q}}{2 m_p}
+
g_A({\vec q}^{~2})\vec{\sigma}
-g_P({\vec q}^{~2})\frac{\vec{q}~ {\vec{\sigma}}_n \cdot \vec{q}}{2 m_p},
\label{a4}
\end{equation}
${\vec r}_n$ is the coordinate of the $n$th nucleon,  $k=1,2,3$, and $g^{\rho,\alpha}$
is the metric tensor.

The two current operators in $M^{0\nu}$ lead to an expression in terms
of 5 parts \cite{Sim99}:
\begin{equation}
M^{0\nu} = M_{VV} + M_{MM} + M_{AA} + M_{AP} + M_{PP},
\label{a5}
\end{equation}
with the notation indicating which parts (axial, vector, etc.) of the nucleon current contribute.
After integrating over $d\vec{x}_1$, $d\vec{x}_2$ and $d\Omega_{q}$ in (\ref{a2})
and writing one-body charge-changing operators in
second quantization as
\begin{equation}
{\hat{\cal O}}_{JM} = \sum_{pn} \frac{\langle p \parallel {\cal O}_J \parallel n\rangle}
{\sqrt{2J+1}} [c^+_p \tilde{c}_n]_{JM},
\label{a6}
\end{equation}
we obtain
\begin{eqnarray}
M_K = \sum_{J,{\pi},k_i,k_f} \sum_{pnp'n'} (-)^J\qquad\qquad\qquad\qquad
\qquad\qquad\\
\frac{R}{g_A^2} \int_0^\infty
\frac{ {\cal P}^K_{pnp'n',J}(q)}
{|q|(|q| + (\Omega^{k_i}_{J^\pi}+\Omega^{k_f}_{J^\pi})/2)} h_K(q^2) q^2 dq
\times
\nonumber \\
\langle 0_f^+ ||
[ \widetilde{c_{p'}^+ \tilde{c}_{n'}}]_J || J^{\pi} k_f \rangle
\langle  J^{\pi} k_f |  J^{\pi} k_i \rangle
 \langle  J^{\pi} k_i || [c_p^+ \tilde{c}_n]_J || 0_i^+ \rangle.
\nonumber
\label{a7}
\end{eqnarray}
Here $K=VV,~MM,~AA,~PP,~AP$
and
\begin{eqnarray}
h_{VV}({\vec q}^{~2} ) &=& - g^2_{V} ({\vec q}^{~2}),\qquad
h_{MM}({\vec q}^{~2}) = \frac{g^2_M ({\vec q}^{~2}) {\vec q}^{~2} }{4m^2_p},\nonumber\\
h_{AA}({\vec q}^{~2}) &=& g^2_A ({\vec q}^{~2}),\qquad\quad
h_{PP}({\vec q}^{~2}) = \frac{g^2_P ({\vec q}^{~2}) {\vec q}^{~4}}{4 m^2_p},
\nonumber\\
h_{AP}({\vec q}^{~2}) &=& -2 \frac{g_A ({\vec q}^{~2})
g_P ({\vec q}^{~2}) {\vec q}^{~2}}{2 m_p}.
\label{a8}
\end{eqnarray}

The reduced matrix elements of the one-body operators
$c_p^+ \tilde{c}_n$ (the tilde denotes a time-reversed state)
in  Eq. (\ref{a7})
depend on the BCS coefficients $u_i,v_j$ and on the QRPA vectors
$X,Y$ \cite{Sim99}. 
%The difference between the QRPA and RQRPA is
%in the way these reduced matrix elements are calculated
%and in the values of RPA energies $\Omega^{k_i}_{J^\pi}$ and
%$\Omega^{k_f}_{J^\pi}$ calculated from the ground
%states of initial $(i)$ and final $(f)$ nuclei. 
The nuclear structure
information resides in these quantities.

The ${\cal P}^K_{pnp'n',J}(q)$ in Eq.\ (\ref{a7})
are products of the reduced one-body matrix elements
of operators ${\cal O}^{(n)}(q)$: 
\begin{eqnarray}
{\cal P}^{VV}_{pnp'n',J}(q) &=&
{\langle p \parallel {\cal O}^{(1)}_J(q ) \parallel n\rangle}
{\langle p' \parallel {\cal O}^{(1)}_J(q) \parallel n'\rangle},
\nonumber\\
{\cal P}^{AA}_{pnp'n',J}(q) &=& \sum_{L=J,J\pm1} (-)^{J+L+1}\times
\nonumber\\
&&{\langle p \parallel {\cal O}^{(2)}_{LJ}(q ) \parallel n\rangle}
{\langle p' \parallel {\cal O}^{(2)}_{LJ}(q) \parallel n'\rangle},
\nonumber\\
{\cal P}^{PP}_{pnp'n',J}(q) &=&
{{\langle p \parallel {\cal O}^{(3)}_J(q ) \parallel n\rangle}
{\langle p' \parallel {\cal O}^{(3)}_J(q) \parallel n'\rangle}},
\nonumber\\
{\cal P}^{AP}_{pnp'n',J}(q) &=& {\cal P}^{PP}_{pnp'n',J}(q),
\nonumber\\
{\cal P}^{MM}_{pnp'n',J}(q) &=& {\cal P}^{AA}_{pnp'n',J}(q)
-{\cal P}^{PP}_{pnp'n',J}(q) ~.
\nonumber\\
\label{a9}
\end{eqnarray}
Here
\begin{eqnarray}
{\cal O}^{(1)}_{JM} (q) &=& 2\sqrt{2} j_J(qr) Y_{JM}(\Omega_r),
\nonumber\\
{\cal O}^{(2)}_{LJM} (q) &=& 2\sqrt{2} j_L(qr)
\{Y_{L}(\Omega_r)\otimes \sigma_1\}_{JM},
\nonumber\\
{\cal O}^{(3)}_{LJM} (q) &=& 2\sqrt{2} \sqrt{\frac{2J-1}{2J+1}}
j_{J-1}(qr) C^{J0}_{J-1 0~10}\times
\nonumber\\
&&~~~~~\{Y_{J-1}(\Omega_r)\otimes \sigma_1\}_{JM}
\nonumber\\
&& -2\sqrt{2} \sqrt{\frac{2J+3}{2J+1}}
j_{J+1}(qr) C^{J0}_{J+1 0~10}\times
\nonumber\\
&&~~~~~\{Y_{J+1}(\Omega_r)\otimes \sigma_1\}_{JM}.
\label{a10}
\end{eqnarray}

%From here one can follow two alternative paths.
%In the first one one would integrate first over $\vec{r}_1$ and  $\vec{r}_2$
%in the one-body matrix elements in the
%Eq. (\ref{a9}) thus obtaining expressions for ${\cal P}^K_{pnp'n',J}(q)$
%that depend only on $q$. The final $M_K$ are then obtained by performing
%the integral over $q$ in Eq.(\ref{a7}). This procedure makes it possible to
%use  single particle wave functions for realistic mean
%field potential instead of commonly used harmonic oscillator ones.
%However, it is impossible (or difficult) to include
%the effect of two-nucleon short range correlations.

The final step, leading to Eq.\ (\ref{eq:long}) in the text, is
to rewrite the product of two one-body matrix elements as
an appropriately recoupled  (with pairs of
protons and neutrons coupled to angular momentum ${\cal J}$)
unsymmetrized two-body matrix element.
Without the complications of angular momentum, this step simply reads
\begin{equation}
 \langle p|O(1)|n \rangle \langle p'|O(2)|n' \rangle =
\langle p, p'| O'(1,2) | n, n'\rangle.
\label{a11}
\end{equation}
We then transform to relative and center-of-mass coordinates
$\vec{r}_{12} = \vec{r}_1 - \vec{r}_2$ and $\vec{R}_{12} = (\vec{r}_1 + \vec{r}_2)/2$.
Since the exchange potential depends only on
$r_{12}=|\vec{r}_{12}|$ we end up with
 Eq.\ (\ref{eq:long}).
The Fermi (F),
Gamow-Teller (GT) and Tensor (T) operators in that equation are
\begin{eqnarray}
O_F(r_{12},E^k_{J^\pi}) &=& \tau^+(1)\tau^+(2) H_F(r_{12},E^k_{J^\pi})\,,
\\
O_{GT}(r_{12},E^k_{J^\pi}) &=& \tau^+(1)\tau^+(2)
H_{GT}(r_{12},E^k_{J^\pi})\sigma_{12}\,,
\nonumber\\
O_{T}(r_{12},E^k_{J^\pi}) &=& \tau^+(1)\tau^+(2)
H_{T}(r_{12},E^k_{J^\pi})S_{12}\,.\nonumber
\label{a12}
\end{eqnarray}
Here
\begin{eqnarray}
\sigma_{12} & = & {\vec{ \sigma}}_1\cdot {\vec{ \sigma}}_2~,~\nonumber \\
S_{12} &=& 3({\vec{ \sigma}}_1\cdot \hat{{ r}}_{12})
       ({\vec{\sigma}}_2 \cdot \hat{{ r}}_{12})
      - \sigma_{12}~.
\label{a13}
\end{eqnarray}
The functions $h_K(q^2)$ that determine the $H_K$'s through the integrals over $q$ in Eq.\ (\ref{eq:pot}) are 
\begin{eqnarray}
	h_F(\vec{q}^{~2}) &=& - g^2_A h_{VV}(\vec{q}^{~2}), \nonumber\\
	h_{GT}(\vec{q}^{~2}) &=&\frac{1}{3}\left( 2 h_{MM}(\vec{q}^{~2})+
	h_{PP}(\vec{q}^{~2})+h_{AP}(\vec{q}^{~2})\right)
\nonumber\\
&& + h_{AA}(\vec{q}^{~2}),
\nonumber\\
h_{T}(\vec{q}^{~2}) &=&
\frac{1}{3}\left(h_{MM}(\vec{q}^{~2})-h_{PP}(\vec{q}^{~2})-h_{AP}(\vec{q}^{~2})\right)\,,
\nonumber\\
\label{a14}
\end{eqnarray}
and the full matrix element is
\begin{equation}
M^{0\nu} = -\frac{M_F}{g^2_A} + M_{GT} + M_T~.
\label{a15}
\end{equation}
Short range repulsion can then be included as explained in Section IV.

\section{} Here we show how to calculate shell-model particle-hole
decompositions so they can be compared with those calculated in the
QRPA. To avoid too many complications, we will use the closure
approximation.
(In the text we have shown that, within the QRPA at least,
using the closure approximation for the $0\nu\beta\beta$-decay
results in an error of $\le$ 10\%.) The matrix element $M$ (the subscript $K$ is implied)
can be written as in Eq.\ (\ref{eq:long}), with the overlap between
intermediate-nucleus eigenstates a Kronecker delta if the those
states are determined uniquely (as in the shell model). The matrix
element $M$ can be decomposed:
\begin{equation}
\label{eq:def} M=\sum_{\alpha} M^{\alpha} \,,
\end{equation}
where $\alpha$ stands for the set of indices $p,p',n,n'$ and
\begin{equation} M^\alpha=\sum_{\mathcal{J}J} s_{\mathcal{J}J}^\alpha
    O_{J}^\alpha \,.
\end{equation}
The parity index $\pi$ is implicitly included along with $J$ and
$\mathcal{J}$.  The $O_{J}^\alpha$ are given by
%\begin{align}
\begin{eqnarray}
 O_{J}^\alpha & = & \sum_{k_i,k_f}  \langle 0_f^+ ||
[ \widetilde{c_{p'}^+ \tilde{c}_{n'}}]_J || J^{\pi} k_f \rangle
\langle  J^{\pi} k_f |  J^{\pi} k_i \rangle \nonumber \\
 &  & \times \langle  J^{\pi} k_i || [c_p^+ \tilde{c}_n]_J || 0_i^+ \rangle ~,
%\end{align}
\end{eqnarray}
and the $s_{\mathcal{J}J}^\alpha$ are everything else in Eq.\ (\ref{eq:long}):
%\begin{align}
\begin{eqnarray}
s_{\mathcal{J}J}^\alpha  =
 (-1)^{j_n + j_{p'} + J + {\mathcal J}}
 \hat{\mathcal J}
\left\{
\begin{array}{c c c}
j_p & j_n & J \\
 j_{n'} & j_{p'} & {\mathcal J}
\end{array}
\right\} Z^\alpha_\mathcal{J}\,,
%\end{align}
\end{eqnarray}
with $\hat{\mathcal{J}} \equiv \sqrt{2\mathcal{J}+1}$ and
\begin{equation}
Z^\alpha_J \equiv \langle p(1), p'(2); {\mathcal J} \parallel
\bar{f}(r_{12}) O_K \bar{f}(r_{12}) \parallel n(1), n'(2); {\mathcal J} \rangle
\,.
\end{equation}
(The $\bar{f}(r_{12})$ can be omitted if short-range correlations are
included some other way.)

 Now we can write $M^{\alpha}$ in two
different ways:
\begin{equation} M^{\alpha} =\sum_\mathcal{J} M^{pp,\alpha}_\mathcal{J} =\sum_J
    M^{ph,\alpha}_J
\end{equation}
with
\begin{equation} \label{eq1} M^{pp,\alpha}_\mathcal{J}=\sum_{J} s_{\mathcal{J}J}^\alpha O_{J}^\alpha
     \,,
\end{equation}
and
\begin{equation} M^{ph,\alpha}_{J}=\sum_\mathcal{J} s_{\mathcal{J}J}^\alpha O_{J}^\alpha\,,
\end{equation}
The $M^{pp,\alpha}_\mathcal{J}$ are the pp-hh amplitudes and the
$M^{ph,\alpha}_J$ are the ph multipole-multipole amplitudes that we
want to calculate in the shell model. All the nuclear structure
information is in the $O_J^\alpha$.

From Eq.\ (\ref{eq1}) we have
\begin{equation} \label{eq2} O_J^\alpha=\sum_\mathcal{J} s^{-1,\alpha}_{J\mathcal{J}}
 M^{pp,\alpha}_\mathcal{J}\,,
\end{equation}
and
%\begin{align}
\begin{eqnarray}
 M^{ph,\alpha}_{J}& = &\sum_\mathcal{J} s_{\mathcal{J}J}^\alpha O_J^\alpha
\\
 && \stackrel{\textrm{(from Eq.\
(\ref{eq2}))}}{-\!\!\!-\!\!\!-\!\!\!-\!\!\!-\!\!\!-\!\!\!\longrightarrow}
\quad \sum_\mathcal{J} s_{\mathcal{J}J}^\alpha
\sum_{\mathcal{J}'}s^{-1,\alpha}_{J\mathcal{J}'}
M^{pp,\alpha}_\mathcal{J'}\,.\nonumber
%\end{align}
\end{eqnarray}
 So, exchanging the primed
and unprimed labels in the sum, we can write the $M^{ph,\alpha}_{J}$
in terms of the $M^{pp,\alpha}_\mathcal{J}$ as
\begin{equation}
\label{eq:phpp} M^{ph,\alpha}_{J} = \sum_{\mathcal{J},\mathcal{J}'}
 s_{\mathcal{J}'J}^\alpha s^{-1,\alpha}_{J\mathcal{J}}  M^{pp,\alpha}_\mathcal{J}
 \,.
 \end{equation}
The final particle-hole multipole contribution that we want is then
just
\begin{equation} M^{ph}_J \equiv \sum_\alpha M^{ph,\alpha}_J \,.
\end{equation}
From the relation
\begin{equation}
\hat{J}^2 \sum_X \hat{X}^2
         \left\{
                       \begin{array}{c c c}
                        a  & b & J \\
                        c & d & X
                       \end{array}
                       \right\}
         \left\{
                       \begin{array}{c c c}
                        a  & b & J' \\
                        c & d & X
                       \end{array}
                       \right\}
               = \delta_{J,J'}
\end{equation}
we have
\begin{equation}
    s^{-1,\alpha}_{J\mathcal{J}} = (-1)^{j_n+j_{p'}+J+\mathcal{J}}
    \frac{\hat{J}^2 \hat{\mathcal{J}}}{Z^\alpha_\mathcal{J}} \left\{
                \begin{array}{c c c}
                   j_p & j_n & J \\
                       j_{n'} & j_{p'} & {\mathcal J}
                \end{array}
                \right\}  \,,
\end{equation}
Finally, putting everything together, we get
%\begin{align}
\begin{eqnarray}
M^{ph}_J & \equiv & \sum_{p,p',n,n',\mathcal{J}\mathcal{J}'}
(-1)^{\mathcal{J}+\mathcal{J}'}\hat{\mathcal{J}}\hat{\mathcal{J}'}
\hat{J}^2 \left\{
        \begin{array}{c c c}
            j_p & j_n & J \\
                j_{n'} & j_{p'} & {\mathcal J}'
             \end{array}
      \right\} \nonumber\\
& & \times \left\{
        \begin{array}{c c c}
            j_p & j_n & J \\
                j_{n'} & j_{p'} & {\mathcal J}
             \end{array}
      \right\}
\frac{Z^\alpha_{\mathcal{J}'}}{Z^\alpha_\mathcal{J}}
M^{pp,\alpha}_\mathcal{J}\,.
%\end{align}
\end{eqnarray}

In a shell model calculation, one can write the double beta-decay
matrix element solely in terms of antisymmetrized matrix elements of
the corresponding operator.  But to obtain the particle-hole
decomposition above, the natural definition since the operator
really represents a product of two one-body currents, one must start
from a representation of the operator in terms of
\emph{unsymmetrized} matrix elements $Z^\alpha_\mathcal{J}$
\begin{equation}
    \hat{O}_K=-\frac{1}{2}\sum_{p,n,p',n',\mathcal{J}}
    Z^\alpha_\mathcal{J} \left[[a^\dag_{j_p}
    a^\dag_{j_{p'}}]^\mathcal{J}  [\tilde{a}_{j_n}\tilde{a}_{j_{n'}}]^\mathcal{J}
\right]^0 \,,
\end{equation}
and calculate the $M^{pp,\alpha}_\mathcal{J}$, for all $\alpha
\equiv p,p',n,n'$, not just $p \geq p', \ n \geq n'$.

%\clearpage

\end{document}